\documentclass[prl,twocolumn,showpacs,superscriptaddress,floatfix]{revtex4}
\usepackage{graphicx}

\def\nbZ{{\mathchoice {\hbox{$\sf\textstyle Z\kern-0.4em Z$}} 
{\hbox{$\sf\textstyle
Z\kern-0.4em Z$}} {\hbox{$\sf\scriptstyle Z\kern-0.3em Z$}} 
{\hbox{$\sf\scriptscriptstyle
Z\kern-0.2em Z$}}}}

\begin{document}

\title{Entanglement in a second order quantum phase transition}
\author{Julien Vidal}
\email{vidal@gps.jussieu.fr}
\author{Guillaume Palacios}
\email{palacios@gps.jussieu.fr}
\author{R\'emy Mosseri}
\email{mosseri@gps.jussieu.fr}

\affiliation{Groupe de Physique des Solides, CNRS UMR 7588, 
Universit\'{e}s Paris 6 et  Paris 7, 2, place Jussieu, 75251 Paris Cedex 05 France}

\begin{abstract}

We consider a system of mutually interacting spin 1/2 embedded in a transverse magnetic field which
undergo a  second order quantum phase transition. We analyze the entanglement properties and the spin
squeezing of the ground state and show that, contrarily to the one-dimensional case, a cusp-like
singularity appears at the critical point $\lambda_c$, in the thermodynamical limit. We also show that
there exists a value $\lambda_0 \geq \lambda_c$ above which the ground state is not spin squeezed despite
a nonvanishing concurrence.

\end{abstract}

\maketitle

Entanglement is a truly specific property of the quantum world, and one of
its deepest signature, as it was already recognized in the early days of quantum
mechanics. It is at the heart of the celebrated EPR paradox \cite{EPR,Bell}, 
and plays a key role in the measurement problem as well as in the quantum to classical
transition \cite{Zurek}. Entanglement is also central in quantum computation
\cite{Eckert_book,Nielsen_book} where the most interesting operations cannot be completely
fulfilled through the manipulation of separable states. 

Recently, entanglement properties of systems undergoing quantum phase transitions
\cite{Sachdev_book} have attracted much attention\cite{Bose,Osborne,Osterloh,Latorre1,Latorre2}.
Interestingly, the concurrence of the ground state which is related to the entanglement of formation
\cite{Wootters98}, has been shown to be strongly affected at the critical point
\cite{Osborne,Osterloh}. More precisely, in the one-dimensional $(1D)$ Ising model in a transverse
field, Osterloh {\it et al.} have shown that the derivative of the concurrence with
respect to the coupling constant diverges at the transition point \cite{Osterloh} although the
concurrence itself is not maximum.  
These pionneering results raise the question of the universality of
these behaviors. Apart from $1D$ quantum spin models, there has been, up to now, no
other analysis of the ground state entanglement in systems displaying quantum phase transitions except in the
Kagom\'e lattice \cite{Bose}.  Actually, the lack of exact solutions especially in higher dimensions implies a 
numerical treatment which often restrict the study to a small number of degrees of freedom. 
Such approaches do not allow, in general, an accurate description of the thermodynamical properties.

In this Letter, we study the entanglement properties of a quantum system made up of $N$ spins
$1/2$ on a simplex (each spin interacts with all others)  embedded in a magnetic field. The
permutation symmetry of this system allows us to restrict the ground state determination to a
$N$-dimensional subspace and,  consequently, to deal with a large number of spins (about one thousand).
We analyze the concurrence and the spin squeezing of the ground state which are, in this case, closely 
related \cite{Wang-Squeezing}. Contrarily to what happens in the $1D$ Ising model, the concurrence of the
ground state is maximum and displays a cusp-like singularity at the critical point. Moreover, at the
transition point, the ground state is maximally spin squeezed and its squeezing parameter
\cite{Kitagawa} vanishes in the thermodynamical limit. Finally, we show that there exists a special line
in the parameter space where the concurrence vanishes and above which the ground state is not spin
squeezed although the concurrence is nonzero.

Let us consider the following Hamiltonian first introduced by Lipkin {\it et al.} \cite{Lipkin}:
%
%
%%%%%%%%%%%%%%%%%%%%
\begin{eqnarray}
H&=&-\frac{\lambda}{N}\sum_{i<j} 
\left( \sigma_{x}^{i}\sigma_{x}^{j} +\gamma\sigma_{y}^{i}\sigma_{y}^{j} \right) 
 -\sum_{i}\sigma_{z}^{i}, \\
&=&-\frac{2 \lambda}{N} \left(S_x^2 +\gamma S_y^2 \right) -2 S_z + {\lambda \over 2} (1+\gamma),
\end{eqnarray}
%%%%%%%%%%%%%%%%%%%%
%
%
where the $\sigma_{\alpha}$'s are the Pauli matrices and  
$S_{\alpha}  =\sum_{i} \sigma_{\alpha}^{i}/2$. We focus here on the ferromagnetic case 
($\lambda>0$) and we mainly consider the case $0 \leq \gamma \leq 1$.  The prefactor $1/N$ is necessary
to get a finite free energy per spin in the  thermodynamical limit. 

The Hamiltonian $H$ preserves the total spin and does not couple states having a different parity of
the number of spin pointing in the magnetic field direction, namely:
%
%
%%%%%%%%%%%%%%%%%%%%
\begin{eqnarray}
[H,{\bf S}^2]&=&0,\\
\left[H,\prod_i \sigma_z^i \right] &=&0.
\end{eqnarray}
%%%%%%%%%%%%%%%%%%%%
%
%
for all $\gamma$. In the isotropic case $\gamma=1$, one further has $[H,S_z]=0$ so that $H$ is 
diagonal in the standard eigenbasis \{$|S,M \rangle$\} of ${\bf S}^2$ and $S_z$. 

For any $\gamma$, this system displays a second order quantum phase transition at 
$\lambda_c=1$ which is characterized by the mean-field exponents \cite{Botetprl,Botetprb}.
Nevertheless, a mean-field approach cannot provide nontrivial entanglement properties since it
essentially turns the Hamiltonian into a sum of single-body Hamiltonians. It is thus necessary to use
numerical diagonalizations of $H$ for finite $N$. 
The dimension of the Hilbert space is $2^N$ but the study of the ground state reduces to a problem
linear with $N$ since it lies in the fully symmetric representation corresponding to the maximum
total spin $S=N/2$. In this subspace spanned by the Dicke states \cite{Dicke}  $|M\rangle=|N/2,M
\rangle$ with 
$M=-N/2,\cdots,+N/2$, one has:
%
%
%%%%%%%%%%%%%%%%%%%%
\begin{eqnarray}
H |M \rangle&=&\left[-{\lambda \over N}(1+\gamma)\left(N^2/4-M^2\right)  -2 M \right]  |M\rangle \nonumber\\
&& 
-
\left(
a_{M-1}^{-} a_{M}^{-} |M-2\rangle  + 
a_{M+1}^{+} a_{M}^{+} |M+2\rangle 
\right) \nonumber\\
&&\times {\lambda (1-\gamma) \over 2 N}
\label{recursion}
\end{eqnarray}
%%%%%%%%%%%%%%%%%%%%
%
%
where 
$a_M^\pm=\sqrt{(N/2)(N/2+1)-M(M \pm 1)}$.
In the following, we will denote by $\cal{E}_\pm$ the orthogonal subspaces spanned by the Dicke states
$|M\rangle$, such that $\prod_i \sigma_z^i |M\rangle=\pm|M\rangle$ which corresponds to even or odd
values of $(N/2-M)$. 

When $\lambda<\lambda_c$ and for any $\gamma$, the ground
state is nondegenerate. By contrast, for $\lambda>\lambda_c$, the ground state is doubly degenerate  
in the thermodynamical limit for any $\gamma\neq 1$ but remains unique in the isotropic case ($\gamma=1$).
In this limit, the magnetization (per spin) in the $z$ direction of the ground state  is simply given by
\cite{Botetprl,Botetprb}:
%
%
%%%%%%%%%%%%%%%%%%%%
\begin{eqnarray}
{1\over N} \langle S_z \rangle &=&{1\over 2} \hspace{10pt} \mbox{for} \:\:\:\: \lambda \leq \lambda_c,\\
{1\over N} \langle S_z \rangle &=&{1\over 2 \lambda} \:\:\:\:\mbox{for} \:\:\:\:\lambda >\lambda_c,
\end{eqnarray}
%%%%%%%%%%%%%%%%%%%%
%
%
for all $\gamma$. 

To analyze the entanglement properties of the ground state $|\psi\rangle$, we have
computed for several values of $\gamma$, the concurrence introduced by Wootters \cite{Wootters98}
which, is defined as follows. Let us denote by $\rho$ the reduced density matrix obtained from
$|\psi\rangle$ by tracing out over $(N-2)$ spins. Of course, in our system,  the choice of the
two spins kept is irrelevant contrarily to the $1D$ Ising model \cite{Osborne,Osterloh}.
Next,  we introduce the spin-flipped matrix 
$\tilde \rho=\sigma_{y}\otimes\sigma_{y}\:\rho^{\ast}\:\sigma_{y}\otimes\sigma_{y}$ where 
$\rho^{\ast}$ is the complex conjugate of $\rho$. The concurrence $C$ is then defined by:
%
%
%%%%%%%%%%%%%%%%%%%%
\begin{equation}
C=\max\left\{ 0,\mu_{1}-\mu_{2}-\mu_{3}-\mu_{4}\right\},
\end{equation}
%%%%%%%%%%%%%%%%%%%%
%
%
where the $\mu_{j}$ are the square roots of the four real eigenvalues, classified in decreasing order,
of the non Hermitian product matrix $\rho \tilde{\rho}$. This concurrence vanishes for an
unentangled two-body state whereas $C=1$ for a maximally entangled one. 
Finally, since $H$ couples every spin with each other, the  two-body  entanglement is somewhat
``diluted" between all spins, and eventually goes to zero in the thermodynamical limit.  
To get nontrivial informations about the entanglement, it is thus crucial to consider the rescaled
concurrence $C_R=(N-1) C$ where the prefactor is simply the coordination number of each spin. In
symmetric multi-qubit systems, this rescaled concurrence has recently been related to the spin squeezing
parameter \cite{Wang-Squeezing}
%
%
%%%%%%%%%%%%%%%%%%%%
\begin{equation}
\xi^2= {4(\Delta S_{\vec{n}_\perp})^2 \over N},
\label{spinsqueezingdef}
\end{equation}
%%%%%%%%%%%%%%%%%%%%
%
%
which measures the spin fluctuations in a correlated quantum state \cite{Kitagawa}.  
The subscript $\vec{n}_\perp$ refers to an axis perpendicular to the mean spin $\langle \vec{S} \rangle$ where
the minimal value of the variance is obtained. More precisely,
for any state belonging to $\cal{E}_+$ or $\cal{E}_-$, one has:
%
%
%%%%%%%%%%%%%%%%%%%%
\begin{equation}
\xi^2=1-C_R,
\label{spinsqueezing}
\end{equation}
%%%%%%%%%%%%%%%%%%%%
%
%
if the matrix elements of the reduced density matrix $\rho$ written in the standard basis
$\{|\uparrow \uparrow \rangle, |\uparrow \downarrow \rangle,|\downarrow \uparrow \rangle,
|\downarrow \downarrow\rangle\}$ satifies: 
$|\rho_{14}|\geq \rho_{22}$ \cite{Messikh,Wang-Squeezing}. In the opposite case, the states are not
spin squeezed ($\xi^2=1$). 

Let us first recall the results in the isotropic case $\gamma=1$ which is exactly solvable. 
As it can be straightforwardly obtained from Eq. (\ref{recursion}), the (nondegenerate) ground
state is the  Dicke state $|N/2 \rangle$ for $\lambda<\lambda_c$, and switches from one state $|M\rangle$
to a state $|M'<M \rangle$ as $\lambda$ increases \cite{Botetprb}. 
The concurrence of a Dicke state $|M\rangle$, can be determined
analytically \cite{Wang-Pairwise,Stockton}: 
%
%
%%%%%%%%%%%%%%%%%%%%
\begin{eqnarray}
C_R&=& \frac{1}{2N} \left\{ N^{2}-4M^{2}- \right.\\
&&\left. \sqrt{( N^{2}-4M^{2}) [(N-2)^{2}-4M^{2}]} \right\}  \nonumber.
\end{eqnarray}
%%%%%%%%%%%%%%%%%%%%
%
%
In the thermodynamical limit, this rescaled concurrence vanishes for $\lambda<\lambda_c$, jumps to $2$ at
the critical point $\lambda_c$, and decreases, by discrete steps, to 1 at large $\lambda$ . This
singular behaviour at the transition point is similar to the one  obtained in the
$1D$ Ising model \cite{Osterloh,Osborne,Latorre1,Latorre2}, except that in this latter
case, the nearest-neighbour concurrence $C(1)$ is not maximum at the critical point. 
Concerning the spin squeezing, its behavior is trivial since all Dicke states are  not spin squeezed
\cite{Wang-Pairwise}, and no singularity can thus be observed on this quantity at the transition.

For $\gamma\neq 1$, the situation is more complex. Indeed, as mentionned above, the ground state
is doubly degenerate for $\lambda>\lambda_c$ in the thermodynamical limit so that, in this region, we
should, in principle, study the entanglement obtained from the thermal density matrix (at zero
temperature)
%
%
%%%%%%%%%%%%%%%%%%%%
\begin{equation}
\rho_{\rm th.}={1 \over 2}\left(|+\rangle \langle + |  + |-\rangle \langle-|\right),
\end{equation}
%%%%%%%%%%%%%%%%%%%%
%
%
where $|+\rangle$ and $|-\rangle$ are {\em any} two orthogonal ground states.
Of course, for finite $N$, the ground state is nondegenerate and lies, depending on $\lambda$,
either in $\cal{E}_+$ or in $\cal{E}_-$. 
In the thermodynamical limit, the reduced density matrices $\rho\pm$ built from the corresponding ground
state $|\psi_\pm \rangle$ by tracing out over $(N-2)$ spins become identical. Therefore, we
have analyzed the entanglement of the true finite $N$ ground state ($|\psi_+\rangle$ or
$|\psi_-\rangle$).

\begin{figure}[ht]
\includegraphics[width=95mm]{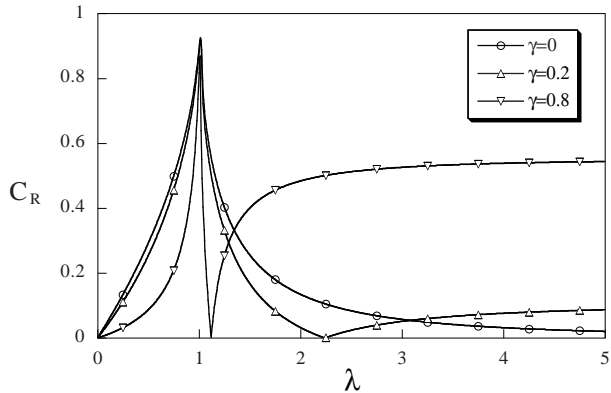}
\vspace{-15mm}
\caption{Rescaled concurrence of the ground state as a function of $\lambda$ for different
values of $\gamma$ and for $N=1000$.}
\label{ground} 
\end{figure}

\begin{figure}[ht]
\includegraphics[width=95mm]{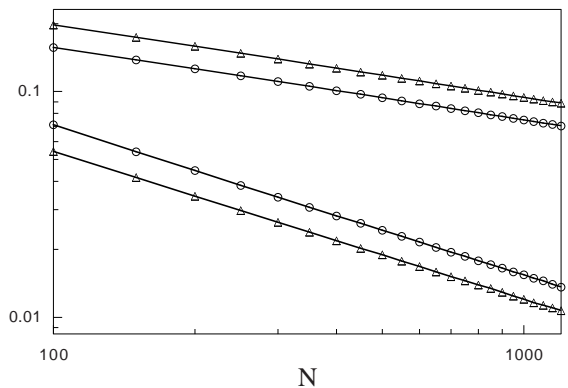}
\vspace{-15mm}
\caption{Behaviors of $1-C_R(\lambda_M)$ (upper curves) and $\lambda_M-1$ (lower curves) as
a function of $N$ for $\gamma=0$ ($\circ$) and $\gamma=1/2$ ($\triangle$).}
\label{scalingC} 
\end{figure}

We have displayed in Fig. \ref{ground}, the rescaled concurrence of the
ground state, as a function of $\lambda$ for various anisotropy parameters 
$0 \leq \gamma\leq 1$.  
%
%
%%%%%%%%%%%%%%%%%%
%%%%%%%%%%%%%%%%%%
%
%
For all $\gamma$, the rescaled concurrence $C_R$ develops a singularity
at the critical point $\lambda_c$ as already pointed out for $\gamma=1$. 
However, as it can be seen in Fig. \ref{scalingC}, the rescaled concurrence goes to
$1$ in the thermodynamical limit contrarily to the isotropic case where it jumps to $2$. More
precisely for all $\gamma\neq 1$, one has: 

%
%
%%%%%%%%%%%%%%%%%%%%
\begin{eqnarray}
1-C_R(\lambda_M)&\sim& N^{-0.33\pm 0.01},\\
\lambda_M-\lambda_c&\sim& N^{-0.66\pm 0.01},
\end{eqnarray}
%%%%%%%%%%%%%%%%%%%%
%
%
where $\lambda_M$ is the value of $\lambda$ for which  $C_R$ is maximum. In the thermodynamical limit,
$C_R(\lambda_M)$ goes to $1$ while $\lambda_M$ goes to $\lambda_c$ so that the ground state is maximally 
spin squeezed at the critical point ($\xi^2=0$).
%
%
%%%%%%%%%%%%%%%%%%
%%%%%%%%%%%%%%%%%%
%
%

\begin{figure}[ht]
\includegraphics[width=95mm]{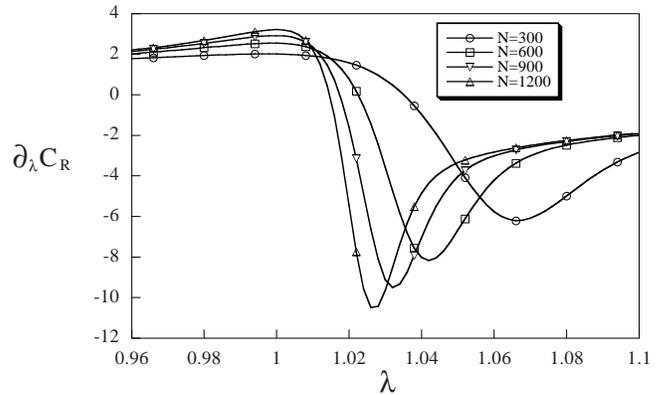}
\vspace{-15mm}
\caption{Finite $N$ behavior of $\partial_\lambda C_R$ for $\gamma=0$ near the critical point
$\lambda_c=1$.}
\label{derivative}
\end{figure}

To analyze the formation of the singularity at $\lambda=\lambda_c$, we have focussed
on the case $\gamma=0$ and plotted in Fig. \ref{derivative}, the behavior of
$\partial_\lambda C_R$ near the critical point,  for different values of $N$. 

%
%
%%%%%%%%%%%%%%%%%%
%%%%%%%%%%%%%%%%%%
%
%

Denoting by $\lambda^{'}_M$ (respectively $\lambda^{'}_m$) the value of $\lambda$ for which
$\partial_\lambda C_R$ is maximum (respectively minimum), one has:
%
%
%%%%%%%%%%%%%%%%%%%%
\begin{eqnarray}
\partial_\lambda C_R(\lambda^{'}_M)&\sim& N^{0.33\pm 0.01},\\
\partial_\lambda C_R(\lambda^{'}_m)&\sim& - N^{0.33\pm 0.03},\\
\lambda_c-\lambda^{'}_M&\sim&N^{-1\pm 0.01},\\
\lambda^{'}_m-\lambda_c&\sim&N^{-0.66\pm 0.01},
\end{eqnarray}
%%%%%%%%%%%%%%%%%%%%
%
%
for all $\gamma \neq 1$ and at large $N$. In the thermodynamical limit, a real cusp-like singularity
is thus observed at the quantum critical point. 
We underline that although we are not able to exactly compute the exponent giving the large $N$ behaviors
of $C_R$, $\partial_\lambda C_R(\lambda^{'}_M)$, and $\partial_\lambda C_R(\lambda^{'}_m)$, we
conjecture that it equals $1/3$. Note that it is also the one guessed in Refs.
\cite{Botetprb,Botetprl} for the scaling of the magnetization at the critical point.

The behaviors of  $C_R$ and $\partial_\lambda C_R$ are notably different from those observed
in the $1D$ case \cite{Osterloh}. Indeed, in the fully connected system considered
here, $C_R$ and $\partial_\lambda C_R$ are extremum at $\lambda_c$ whereas in
the $1D$ Ising model, $\partial_\lambda C(1)$ is the only quantity affected by the
transition ($C(1)$ is surprisingly maximum below the critical point). In addition, the scaling
behavior of the concurrence and of its derivative are different in both models. This simply
reflects the fact that they do not belong to the same universality class as it was already known
from the calculations of critical exponents. 

In the zero coupling limit ($\lambda=0$), the rescaled concurrence obviously vanishes since the ground
state is in this case the fully polarized Dicke state  $|N/2\rangle$ and, accordingly $\xi^2=1$. More
interestingly, for $\gamma \neq 0$, there exists another special value $\lambda_0(\gamma)$ 
for which $C_R$ vanishes. For $\lambda \geq \lambda_0(\gamma)$, the rescaled concurrence is
nonzero but the ground state is not spin squeezed $(\xi^2=1)$, whereas for 
$\lambda < \lambda_0(\gamma)$, the spin squeezing is given by (\ref{spinsqueezing}). This 
behavior of $\xi^2$ is due to a change in the sign of $|\rho_{14}|-\rho_{22}$ which is always 
negative above $\lambda_0(\gamma)$. For $\gamma=0$, such a situation never occurs and the ground
state is always spin squeezed. This is a surprising result since it singularizes the case $\gamma=0$
that, from the phase transition viewpoint, belongs to the same universality class as the case
$\gamma\neq 0$ \cite{Botetprl,Botetprb}.  

%
%
%%%%%%%%%%%%%%%%%%
\begin{figure}[ht]
\includegraphics[width=95mm]{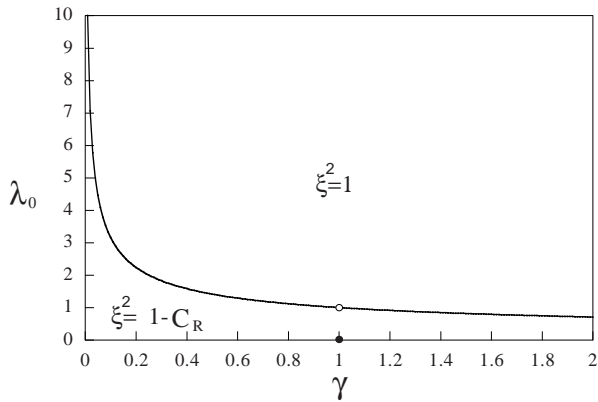}
\vspace{-15mm}
\caption{Phase diagram for the ground state spin squeezing in the plane $(\gamma,\lambda_0)$. Note that
for the isotropic case $\gamma=1$, the ground state which is a Dicke state is never spin squeezed
($\xi^2=1$ for any $\lambda$).}
\label{lambdac}
\end{figure}
%%%%%%%%%%%%%%%%%%
%
%
We have displayed in Fig. \ref{lambdac} the ``critical line" $\lambda_0(\gamma)$.
Apart from the very specific case $\gamma=1$ for which the ground state is never spin squeezed, this
line is given, in the thermodynamical limit, by: $\lambda_0=1/\sqrt{\gamma}$ \cite{Palacios2}. 
We emphasize that this formula is also valid for  $\gamma>1$ though, in this region, the critical point
is readily obtained by a rescaling of the coupling constant and is given by $\lambda_c=1/\gamma$. 

At this stage, we do not completely understand why the entanglement properties are so strongly affected
by the quantum critical point. In particular, the extremization of the (rescaled) concurrence does not
seem to be a generic characteristic since this phenomenon is  not observed in the $1D$ Ising model in a
transverse magnetic field \cite{Osborne,Osterloh}, at least for $C(1)$ \cite{C2}.  Note however that in
both models, the variation of the concurrence is extremal at $\lambda_c$. 
Although in the present case, we have not exactly related the scaling exponent of the entanglement to the
critical exponents, there may certainly exists some deep relations between them which deserves further
investigations. It would also be interesting to analyze the scaling of the Von-Neumann entropy which
has  been, very recently, related to the central charge  of the conformal theory associated to the $1D$
quantum spin models \cite{Latorre1,Latorre2}.

Several important issues remains opened. In other systems displaying a quantum phase transition, the
behavior of the spin squeezing has never been investigated so far. It would be worth determining
whether it is always minimum at the critical point or not. Indeed, if the concurrence is not always
maximum at the transition, nothing prevents the spin squeezing to be minimum as it is the case in the
present study. Another challenging question concerns the quantum dynamics. For nonstationary states, one
may wonder how the proximity of a quantum critical point influences the time evolution of the
entanglement.  For a simple initial state fully polarized along the field direction, we have already
some indications that in the fully connected system analyzed here, the rescaled concurrence vanishes, at
larger times,  for $\lambda\geq\lambda_c$ \cite{Palacios2}. Though we cannot assert that it is a generic
situation, it is likely that the entanglement of all eigenstates is modified at the critical point and
consequently, the one of any quantum states built from them. Such a study would be of primer interest in
exactly solvable models.

\acknowledgments 

We are very grateful to C. Aslangul, C. Caroli and B. Dou\c{c}ot for fruitful and stimulating
discussions.


\begin{thebibliography}{10}
\bibitem{C2}
The next-nearest-neighbour concurrence $C(2)$ is indeed maximum at the critical point but its
first  derivative with respect to $\lambda$ vanishes.
\end{thebibliography}
\end{document}